\input{aipcheck}

\documentclass[
    ,final            % use final for the camera ready runs
  ]
  {aipproc}

\layoutstyle{6x9}

\begin{document}

\title{Inclusive Search for the SM Higgs Boson in the $H\to\gamma\gamma$ channel at the LHC}


\author{Serguei Ganjour}{
  address={CEA-Saclay/IRFU, F-91191 Gif-sur-Yvette, France\\
  {\rm E-mail: Serguei.Ganjour@cea.fr}}
}

\begin{abstract}
 A prospective for the inclusive search of the Standard Model Higgs boson
in the decay channel $H\to\gamma\gamma$ is presented with the CMS
experiment at the LHC. The analysis relies on a strategy to determine the
background characteristics and systematics from data. The strategy is
applied to a Monte Model of the QCD background, with full simulation of
the detector response. The discrimination between signal and background
exploits information on photon isolation and kinematics. The resolution for
the reconstructed Higgs boson mass profits from the excellent energy
resolution of the CMS crystal calorimeter. A discovery significance above 5 sigma
is expected at integrated LHC luminosities below 30~fb$^{-1}$ for Higgs
boson masses below 140~GeV/c$^2$.
\end{abstract}

\maketitle

%%%%%%%%%%%%%%%%%%%%%%%%%%%%%%%%%%%%%%%%%%%%
%% MAINMATTER
%%%%%%%%%%%%%%%%%%%%%%%%%%%%%%%%%%%%%%%%%%%%

\section{Introduction}

Since the beginning of the LHC project the $H\to\gamma\gamma$ channel is considered as 
a major  discovery channel of Higgs particles~\cite{Htogg:proposal} at masses between LEP
limit 114.4~GeV/c$^2$~\cite{LEP:Limit} 
and about 140~GeV/c$^2$. 
Indeed, despite a high rate of the  $H\to b\bar{b}$ decays at low mass, it remains  
out of our interest due to large QCD background and low mass resolution of the di-jet 
state.  Since the $H\to\gamma\gamma$ decay involves virtual loops, it has relatively 
small branching fraction of about 10$^{-3}$.  However, the signal has a clean signature 
with two high E$_T$ isolated photons. Due to an excellent resolution of 
the CMS electromagnetic calorimeter (ECAL), it can be identified as 
a narrow peak at a Higgs boson mass on the top of a continuous background.

We investigated two analysis methods for the $H\to\gamma\gamma$ searches at CMS~\cite{pTDR2,Htogg:CmsNote}. 
In addition to the conventional cut-based analysis, 
we report the optimized discovery oriented technique
based on a multivariate optimization.
The latter exploits a difference in the signal and the background kinematics.
These studies use a full CMS detector simulation program assuming $2\times10^{33}$cm$^{-2}$s$^{-1}$
machine luminosity and including collision effects such as minimum bias and underling events 
in the  simulation model.

\section{Rates and cross sections}

The inclusive search of the Higgs boson implies any production mechanism. 
In proton-proton collisions at LHC energy about 80\% of the Higgs bosons are produced in the gluon fusion reaction, 
while the rest are produced in association with either $q\bar{q}$ pairs (WVB or $t\bar{t}$ fusion)
or vector bosons. Table~\ref{tab:signal-cs} presents the cross sections and $H\to\gamma\gamma$ 
branching fractions for the different Higgs boson mass~\cite{Htogg:Spira}. 

\begin{table}[!htbp]
\begin{tabular}{lrrrrr}
\hline
  M$_{\rm H}$ (GeV/c$^2$)               &  115    &  120  &   130   &  140   & 150  \\ \hline
  $\sigma$ gg fusion (pb)           &  39.2   &  36.4 &   31.6  &  27.7  & 24.5 \\ 
  $\sigma$ WVB fusion (pb)          &  4.7    &  4.5  &   4.1   &  3.8   & 3.6  \\ 
  $\sigma$ WH, ZH, $t\bar{t}$H (pb) &  3.8    &  3.3  &   2.6   &  2.1   & 1.7  \\ 
Total $\sigma$  (pb)                &  47.6   &  44.2 &   38.3  & 33.6   & 29.7 \\
$\mathcal{B}$(H$\to\gamma\gamma$), \%         &  0.21   & 0.22  &   0.22  & 0.20   & 0.14 \\ \hline
$\sigma\times\mathcal{B}$ (fb)              &  99.3   &  97.5 &   86.0  & 65.5   & 41.5 \\ \hline
\end{tabular}
\caption{Signal NLO cross sections and branching ratios.}
\label{tab:signal-cs}
\end{table}

We consider two sorts of the background processes. The ``irreducible'' background has 
two real high E$_T$ isolated photons. Such a signature can be produced by both quark-antiquark annihilation (``born'') 
and gluon-gluon fusion (``box'') as well as quark-gluon Compton scattering with isolated bremsstrahlung processes. We estimate 
the total differential rate of ``irreducible'' backgrounds about 100~fb/GeV/c$^2$ at 120~GeV/c$^2$ mass. 
Thus, we require about 1~GeV/c$^2$ two-photon mass resolution for a powerful discrimination of the signal 
(see Table~\ref{tab:signal-cs}).

\begin{table}
\begin{tabular}{lrr}
\hline
  Process~~~~~~~~~~~~~ & ~~~~~~~~~~~~~~~~~~~~~~$p_T$        &~~~~~~~~~$\sigma_{\rm LO}$ \\ 
         & (GeV/c)      & (pb)            \\ \hline
  pp$\to\gamma\gamma$ (born)  & $>25$ & 82  \\ 
  pp$\to\gamma\gamma$ (box)   & $>25$ & 82  \\ 
  pp$\to\gamma+$jet           & $>30$ & $5\times10^{4}$  \\ 
  pp$\to$jets                 & $>50$ & $2.8\times10^{7}$  \\ 
  Drell Yan ee                & $-$ & $4\times10^{3}$     \\ \hline
\end{tabular}
\caption{LO cross sections for backgrounds.}
\label{tab:bkg-cs}
\end{table}

The dominant QCD processes like $\gamma+$jet and di-jets may lead to the fake photons 
induced by neutral hadrons $\pi^0$ or $\eta$ and produced in the jet fragmentation processes. 
The ``reducible'' background has at least one non-isolated photon.
The PYTHIA LO cross sections of the background processes are presented in Table~\ref{tab:bkg-cs}. 
To compute the background yields at NLO cross sections we apply the K-factors summarized in 
Table~\ref{tab:k-factor}~\cite{Htogg:bkg}.
Thus, a jet suppression has to be better then 10$^{-3}$ to diminish its rate 
at the level of the ``irreducible'' background.

\begin{table}
\begin{tabular}{lr}\hline
  Process & K-factor \\ \hline
  pp$\to\gamma\gamma$ (born)  & 1.5 \\  at the  test-beam.  
  pp$\to\gamma\gamma$ (box)   & 1.2 \\ 
  pp$\to\gamma+$jet (2 prompt)            & 1.72 \\ 
  pp$\to\gamma+$jet (1 prompt)            & 1 \\ 
  pp$\to$jets                 & 1    \\ \hline
\end{tabular}
\caption{Background K-factors applied for the PYTHIA cross section}
\label{tab:k-factor}
\end{table}

\section{Detector}

The conceptual design of the CMS detector ~\cite{pTDR1} exploits 
the conventional layout of the particle detector at hadron colliders.
A dedicated electromagnetic calorimeter is required 
for a powerful search of the Standard Model (SM) Higgs boson.
Indeed, the $H\to\gamma\gamma$ decay was employed as a benchmark 
channel in the design of the ECAL. It comprises
75848 lead tungstate (PWO$_4$) crystals placed in two pseudo-rapidity regions:
the barrel ($|\eta|<1.479$) and the endcap ($1.479<|\eta|<3.0$).
The ECAL endcap is equipped by a preshower (ES) system for $\pi^0$ rejection. 
Due to low Moliere radius (2.19~cm) about 80\% of the shower energy 
is deposited in one crystal having  2.2$\times$2.2~cm$^2$ front face size.
Such a high granularity allows us a powerful identification of isolated photons. 

\begin{figure}[!htbp]
  \includegraphics[width=.8\textwidth]{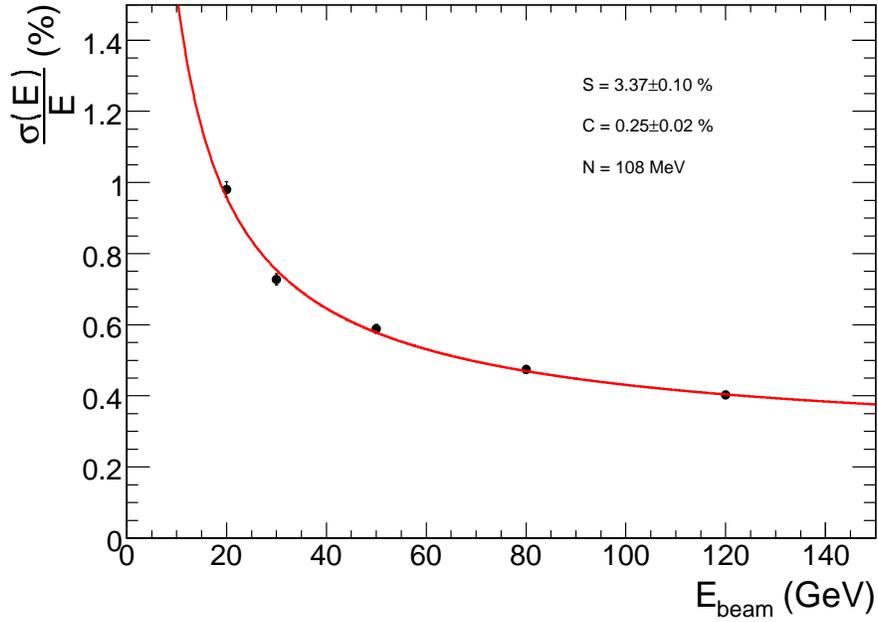}
    \caption{Energy resolution of the barrel supermodule.}
    \label{fig:EBres}
\end{figure}

Figure~\ref{fig:EBres} shows the energy resolution of the ECAL barrel obtained with an electron test-beam. 
The stochastic ($S$), the noise ($N$) and the constant ($C$) terms are obtained by a fit with the following function:
\begin{equation}
\biggl(\frac{\sigma}{E}\biggr)^2=\biggl(\frac{S}{\sqrt{E}}\biggr)^2+\biggl(\frac{N}{E}\biggr)^2+C^2.
\label{math:EB-res}
\end{equation} 
Clustering algorithm accounts strong magnetic field and presence of material in front of the calorimeter. 
Measured energy can be computed as
\begin{equation}
E_{e, \gamma}= F\times \sum_{clusters}Gc_iA_i,
\label{math:ECALcalib}
\end{equation} 
where $F$ is a correction function, $Gc_i$ is a calibration factor 
expressed as a product of a global absolute scale and intercalibration constants, respectively. 
$A_i$ is a signal amplitude, in ADC counts. Sum of the incident energy 
in 5$\times$5 crystals provide the best energy estimation for the unconverted photons ($F=1$) .  
The dedicated procedure of ECAL calibration is foreseen 
with the physics events such as  $W\to e\nu$, $Z\to e^+e^-$ 
and $\pi^0\to\gamma\gamma$.

\section{Analysis and results}

Higgs decays into two-photon state are selected with extremely high efficiency  both Level-1 (99.7\%)
and High Level Triggers (88.4\%). Due to tighter kinematical and isolation critera 
applied in the analysis, no additional signal restriction by trigger were observed.

Two-photon candidates are selected within fiducial volume $|\eta|<2.5$, and $p_T^\gamma>40,\ 35$~GeV/c.
Since fake photons produced in the jets are accompanied by additional energetic particles, 
we require no tracks with $p_T>1.5$~GeV/c inside a cone around photon with a radius $\Delta R<0.3$. 
In addition, we demand $\sum E_T<6(3)$~GeV/c in the ECAL  barrel (endcap) in a cone $0.06<\Delta R<0.35$ 
and  $\sum E_T<6(5)$~GeV/c in the HCAL  barrel (endcap) in a cone  $\Delta R<0.3$. 
The energy response of individual crystals were smeared 
according to calibration precision expected for an integrated 
luminosity of 10~fb$^{-1}$ and obtained with $W\to e\nu$ control sample.

Due to longitudinal spread of the interaction vertices ($\sim50$~mm) 
the di-photon mass is smeared by about 1.5~GeV/c$^2$.
Indeed, Higgs bosons are produced in association with tracks from
underlying events, initial state gluon radiation and associative particles 
$qqH$, $WH$, $ZH$. These tracks allow us to identify the interaction vertex, 
in about 80\% of cases, and correct momenta of the photons.

\begin{figure}[!htbp]
  \includegraphics[width=.8\textwidth]{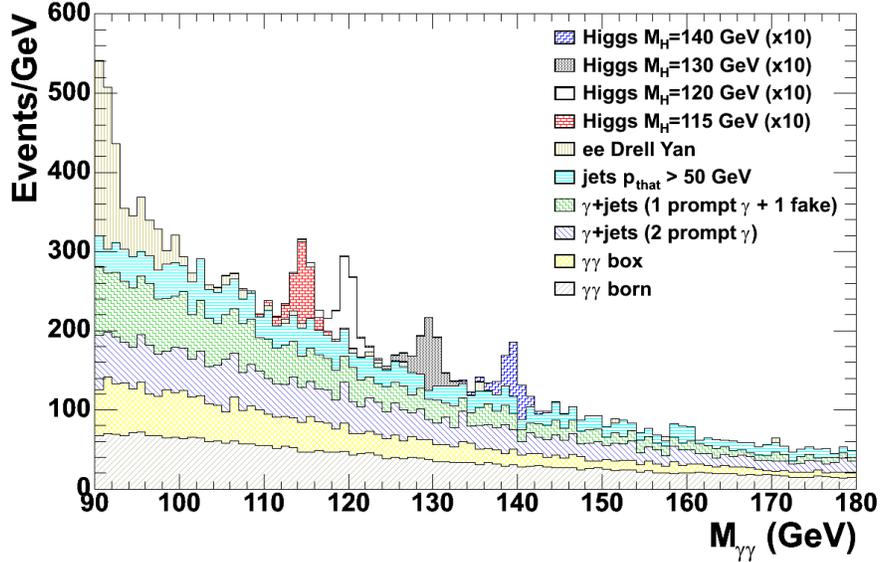}
    \caption{Expected di-photon invariant mass distribution for the cut-based analysis 
    normalized to an integrated luminosity of 1~fb$^{-1}$. Signal is scaled by a factor 10.}
    \label{fig:mgg}
\end{figure}

Figure~\ref{fig:mgg} shows the di-photon mass distribution 
for the signal (scaled by a factor 10) and the background events after 
the selection. For a 120 GeV/c$^2$ Higgs boson, 
the signal efficiency of 30\% yields to 29.3 observed Higgs 
events per inverse femtobarn, while the total background is 178 fb/GeV/c$^2$ 
as shown in Table~\ref{tab:bkg-rates}. Despite asymmetric lineshape about 60\% 
of the signal remains within $\pm$1~GeV/c$^2$ mass window. 

\begin{table}[!htbp]
\begin{tabular}{lrrrr}
\hline
  Process               & \multicolumn{4}{c}{$M_{\rm H}$, (GeV/c$^2$ )}\\ 
                        & 115 & 120 & 130 & 140 \\ \hline
  pp$\to\gamma\gamma$ (born)  &    48                 &    44                 & 36 &    29                  \\
  pp$\to\gamma\gamma$ (box)   &    36                 &    31                 & 23 &    16                  \\
  pp$\to\gamma+$jet (2prompt) &    43                 &    40                 & 32 &    26                  \\          
  pp$\to\gamma+$jet (1prompt) &    40                 &    34                 & 22 &    19                  \\          
  pp$\to$jets                 &    29                 &    27                 & 20 &    18                  \\
  Drell Yan ee                &    2                  &    2                  & 1  &    1                   \\ \hline
  Total background            &   203		      &   178		      & 134&   109 \\ \hline
\end{tabular}
\caption{Background yields in fb/GeV/c$^2$}
\label{tab:bkg-rates}
\end{table}

Including of new variables or cuts optimization do not lead to the further improvement 
of the signal over background ratio ($s/b$). However, we find the R$_9$, 
fraction of the super-cluster energy deposited a small 3$\times$3 crystal area,
as a powerful discriminator of a $\pi^0$. 
Indeed, high value of R$_9$ readily identifies converted photons 
and automatically selects against $\pi^0$. The converted category 
remains background enriched. However, the use of a track information 
would allow further rejection of the background.

To improve signal significance, we split the analyzed sample 
into categories with different $s/b$. Asides {\it single category}, 
we consider {\it four categories}, 2 categories where both photons are detected in the barrel 
and 2 where at least one photon  is in the endcap. Each of  these 2 categories 
are splitted for high ($>$0.93) and low ($<$0.93) values of R$_9$. 
Then, we form {\it twelve categories} sample based on
3 ranges in $R_9$ (0.9, 0.948) and 4 pseudo-rapidity 
regions in $|\eta|$ (0.9, 1.4, 2.1).
Table~\ref{tab:lum-discovery} summarizes the integrated luminosity needed to discover or to exclude 
Higgs boson for the different event splitting. Confidence level is calculated with a frequentistic 
approach using a log-likelihood ratio (LLR) estimator. These outcomes are computed for many possible 
pseudo-experiments for a hypothesis when the signal exists and that it does not.

\begin{table}[!htbp]
\begin{tabular}{lccc}\hline
                        & $5\sigma$        & $3\sigma$        & $95\%$ C.L. excl.\\ \hline
1 category              & 24.5  (39.5)     & 8.9  (11.5)      & 4.1 (5.8) \\
4 categories            & 21.3  (26.0)     & 7.5  (9.1)       & 3.5 (4.8) \\
12 categories           & 19.3  (22.8)     & 7.0  (8.1)       & 3.2 (4.4) \\ \hline
\end{tabular}
\caption{Integrated luminosity required for observation or exclusion of the Higgs boson with a mass of 120~GeV/c$^2$.}
\label{tab:lum-discovery}
\end{table}

The optimized analysis exploits six categories, 3 categories 
where both photons are detected in the barrel and 3 where at least one photon  
is in the endcap. These 3 categories are defined according to measured R$_9$, as for the cut-based analysis 
splitted into 12 categories. Loose cuts are applied for the isolation variables which are used as a 
neural network prior. Fixed cut for the optimized NN$_{iso}$ output variable slightly improves 
a background rejection. Asides two-photon mass, four other kinematical variables can be used for further 
discrimination of the signal. They are transverse energy of each photon $E_T^{\gamma1,2}$, $\eta$ 
difference between two photons $|\eta_1-\eta_2|$ and longitudinal momentum of the photon pair $P_L^{\gamma\gamma}$. 
These variables are combined with NN$_{iso}$ for the further neural network optimization. 
To reject the background rate a neural net is trained with the mass side-band events and results into NN$_{kin}$ output variable.
Due to negligible correlation between invariant mass and NN$_{kin}$, the $s/b$ expectation can be estimated for each event as 
\begin{equation}
\biggl(\frac{s}{b}\biggr)_{est}=\biggl(\frac{s}{b}\biggr)_{mass}\times\biggl(\frac{s}{b}\biggr)_{kin}.
\label{math:LH-opt}
\end{equation}
Finally, the events are binned according to the $s/b$ estimate. 
Then, we exploit $\log(s/b)$ distribution to compute the confidence 
level with the LLR estimator. 

\begin{figure}[!htbp]
  \includegraphics[width=.8\textwidth]{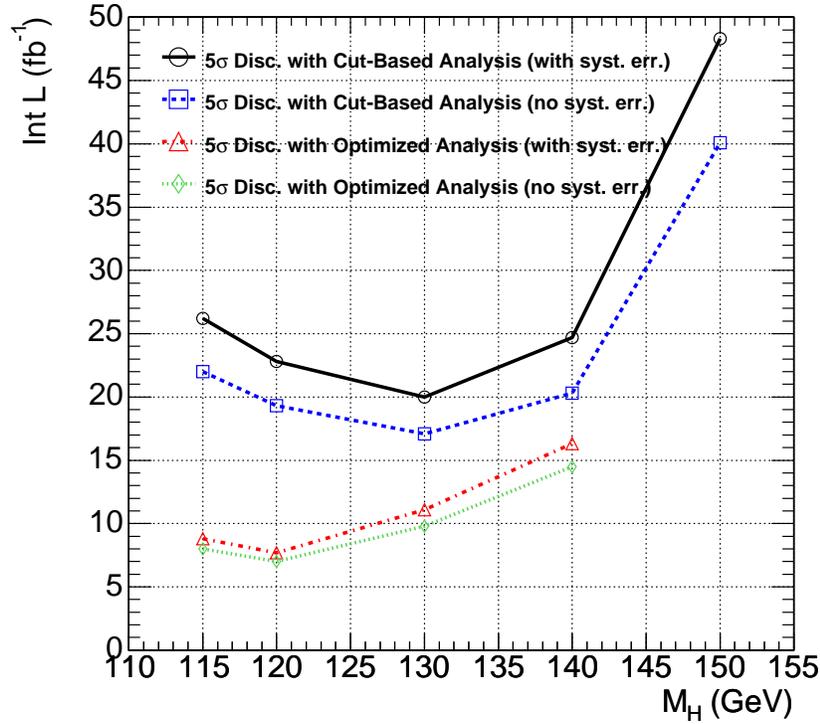}
    \caption{Integrated luminosity required for a 5$\sigma$ discovery as a function of the Higgs mass.}
    \label{fig:5sigma}
\end{figure}

Figure~\ref{fig:5sigma} shows the integrated luminosity needed for 
a 5$\sigma$ discovery. For a 120~GeV/c$^2$ Higgs boson, such a discovery 
can occur with about 7.7~fb$^{-1}$ recorded data using an optimized technique, 
while it is required about  22.8~fb$^{-1}$ of integrated luminosity  
using conventional cut-based analysis technique (see Table~\ref{tab:lum-discovery}). 
We address major systematic errors. The dominant contribution is due to background subtraction. 
It is evaluated from the uncertainty of the fit function in the mass side-bands. 

\section{Conclusions}

The SM Higgs boson can be readily discovered with an integrated luminosity 
less then 30~fb$^{-1}$,  if its mass is below 140~GeV/c$^2$. 
We investigated the standard cut-based and discovery oriented optimized analysis techniques.
The best achievement corresponds to a 120~GeV/c$^2$ Higgs boson
where  7.7~fb$^{-1}$ of a recorded data is required for the 5$\sigma$ discovery. 
The analysis strategy implies data driven methods of the background subtraction
using a side-band mass regions. There is a potential room for improvement 
by adding the track information to identify converted photons and ECAL preshower 
to reject $\pi^0$.

%%%%%%%%%%%%%%%%%%%%%%%%%%%%%%%%%%%%%%%%%%%%%%%%
%% BACKMATTER
%%%%%%%%%%%%%%%%%%%%%%%%%%%%%%%%%%%%%%%%%%%%%%%%

%\begin{theacknowledgments}

%\end{theacknowledgments}



\bibliographystyle{aipproc}   % if natbib is available

%\bibliography{sample}

\IfFileExists{\jobname.bbl}{}
 {\typeout{}
  \typeout{******************************************}
  \typeout{** Please run "bibtex \jobname" to optain}
  \typeout{** the bibliography and then re-run LaTeX}
  \typeout{** twice to fix the references!}
  \typeout{******************************************}
  \typeout{}
 }

%%%%%%%%%%%%%%%%%%%%%%%%%%%%%%%%%%%%%%%%%%%
%% The following lines show an example how to produce a bibliography
%% without the help of the BibTeX program. This could be used instead
%% of the above.
%%%%%%%%%%%%%%%%%%%%%%%%%%%%%%%%%%%%%%%%%%%

\end{document}